\documentclass[letterpaper, 10 pt, conference]{ieeeconf}  
\IEEEoverridecommandlockouts                          

\usepackage{times}
\usepackage{cite}
\usepackage{graphicx, subfigure, epsfig}
\usepackage{url}
\usepackage{algorithm}
\usepackage{algorithmic}
\usepackage{amsmath}
\usepackage{xspace}
\usepackage{capt-of}

\begin{document}

\title{\LARGE \bf
Efficient Sampling for Better OSN Data Provisioning}

\author{Nick Duffield$^{*}$ and Balachander Krishnamurthy$^{\ddagger}$
\thanks{$^{*}$Nick Duffield is with the Departments of Electrical \&
  Computer Engineering and Computer Science \& Engineering, Texas A\&M University, College Station, TX
  77843, USA
        {\tt\small duffieldng@tamu.edu}}%
\thanks{$^{\ddagger}$Balachander Krishnamurthy is with AT\&T Labs Research,
  33 Thomas Street, New York, NY 10007, USA 
        {\tt\small bala@research.att.com}}%
}

\newcommand\E{\textsf{E}}
\def\sts{\textsc{Sts}}
\def\asu{\textsc{Crawl}}
\def\stslong{\textsc{StreamSample}}
\def\us{\_}
\newcommand{\be}{\begin{equation}}
\newcommand{\ee}{\end{equation}}
\newcommand{\bea}{\begin{eqnarray}}
\newcommand{\eea}{\end{eqnarray}}
\let\hat\widehat
\def\down#1{_{(#1)}}
\def\up#1{^{(#1)}}

\def\cre{\textsc{cr}}
\def\uid{\textsc{id}}
\def\sta{\textsc{st}}
\def\act{\textsc{ac}}
\def\fo{\textsc{fo}}
\def\fr{\textsc{fr}}
\def\rep{\textsc{re}}
\def\itr{\textsc{itr}}
\def\fb{\textsc{fb}}
\def\sd{\textsc{sd}}
\def\id{\textsc{id}}
\def\ffan{\textsc{ffan}}
\def\uni{\textsc{uni}}
\def\var{\mathrm{var}}

\maketitle

\begin{abstract}
  Data concerning the users and usage of Online Social Networks (OSNs)
  has become available externally, from public resources (e.g., user
  profiles), participation in OSNs (e.g., establishing relationships
  and recording transactions such as user updates) and APIs of the OSN
  provider (such as the Twitter API). APIs let OSN providers
  monetize the release of data while helping control measurement
  load, e.g. by providing samples with different cost-granularity
  tradeoffs. To date, this approach has been more suited to 
  releasing transactional data, with graphical data still being
obtained by resource intensive methods such a graph crawling. In
  this paper, we propose a method for OSNs to provide samples of the
  user graph of tunable size, in non-intersecting increments, with
  sample selection that can be weighted to enhance accuracy when
  estimating different features of the graph.
\end{abstract}

\section{Introduction}
\label{sec:intro}

Online Social Networks (OSNs) continue to grow rapidly with 1.7
billion monthly active users on Facebook\footnote{Data from Facebook
  for the
  2$^{\textrm nd}$ quarter of 2016} and over 300 million on
Twitter\footnote{Data from Twitter for June 2016}.  OSN providers are
making small portions of their vast data collections available to
different external parties, including researchers, based on business
considerations (increasing reach and sales or enabling external
application writers). The manner of providing access, the choice of
the samples, etc., are all unilateral decisions made by the OSN
providers. In this paper, we argue that there are better ways to make
samples of user data available to external parties that would benefit
all.  We examine current practices in OSN data availability, the
manner of analyses which use the data, and the challenges posed by
scale and statistical features of the data itself. We propose sampling
schemes that accommodate the need estimate accurately in multiple data
dimensions and can also serve different sampling rates without duplication under a controllable database load.

We examine current practices in OSN data availability, the manner of
analyses which use the data, and the challenges posed by scale and
statistical features of the data itself. Externally to OSNs there is
much interest in understanding users, their OSN relationships, and the
OSN events they generate. Research aims to understand
statistical properties, such as connectivity, clustering, node degree,
events. Increasingly, there is commercial interest in identifying
specific subsets of the OSN graph (active users, highly connected and
influential users, groups of users with specific common
interests) and subsets of OSN events, including updates containing
specific keywords, such as a company name, product, or service.
  
\subsection{Current trends and our motivation}

There are currently two types of methods to obtain information concerning
OSN networks. The first is direct measurement of eternally visible
portions of the OSN, possibly as a member, (e.g., viewing accessible portions
of users' profile pages, or registering for content updates). The
second is by employing tools provided by the OSN, such as the Twitter
API, that give partial visibility into users relationships, and the
stream events (status updates, i.e. tweets). 
Much research activity (see Section~\ref{sec:related}) has focused 
both on developing strategies for efficiently using these resources and
estimating OSN properties from the subsets of data so obtained.  This
approach often involves traversal of portions of the OSN graph, en
passant compiling relatively large data subsets for research and
evaluation purposes. However, such approaches are relatively costly in
time, resources, and expertise, and hence not feasible for
non-specialist use. 
  
A more effective model for non-specialist users positions them at the
end of a chain of commercial relationships. The chain starts with OSNs
monetizing access to their user data by providing partial data feeds
with members selected, e.g., by sampling or attribute-based filtering,
potentially at different price points depending on factors such as
the sampling rate and reported detail per item.
Independent OSN analytic services subscribe to these data feeds,
possibly joining them with OSN data measured independently, and other
data sources such as demographic information. This provides downstream
feeds and/or query functionality that can be sold to non-specialist
end users. The OSN analytic services would subscribe to
higher-cost data feeds than would be economical for an individual user,
with the costs effectively being divided over the set of end users.
 
In order to realize the model just described, 
what is needed
is \textsl{a systematic way for OSNs to release samples of graph level
objects---such as users (nodes), relationships between users (links)
and node clusters---with a level of sampling that can be tuned to a
desired price point.} OSNs primarily benefit in two ways. First, this
produces a new data feed that can be monetized. Second, this reduces the
measurement based load on the OSN service network, since it provides
information that would otherwise be obtained only by direct measurement of
the OSN. A corollary of this second point it that 
OSNs have an incentive to employ a sampling strategy that, for a given
data feed sampling rate \textsl{optimizes the accuracy of analyses built on
the feed.} This reduces any incentive for subscribers to revert to
independent measurement.
 
How should the OSN sample its user graph? The constraints for external
measurement do not apply to the OSN provider.  We assume that the OSN
stores the user graph in a database and that it is feasible to pass
exhaustively over all elementary graph objects (nodes and links) in
order to compile samples. Composite objects such as node clusters of a
given size may also be amenable to treatment. 
Within this framework, a sampling
methodology should exhibit the following three properties:

\paragraph*{S1} \textsl{Multidimensional Estimation Accuracy:} There are
  many interesting properties for OSN analytics, including
  node properties (e.g., event activity rate, node in and out
  degree) and link properties (e.g., degree pairs of connected
  nodes).
  The sampling methodology must be able to yield accurate
  estimates of the marginal distributions of the variables, e.g., to
  service queries on the distribution of the number of connection
  beween OSN users.
However for OSN analytics, there is a great interest in
  understanding the \textsl{relationships between user properties}, as
  expressed through queries such as finding the proportion of all connections
  or of all event activity that is represented by a given proportion
  of the most connected users (see e.g. \cite{{twit:huberman}}). Answering such queries requires
  accurate estimation of not only the joint distribution of the
  variables concerned, but the \textsl{joint mass distributions}, i.e.,
  the joint distributions weighted by the values of one of more variables.

\paragraph*{S2} \textsl{Single Pass Serving of Different Sampling Rates:} 
  rather than have separate database passes for each sampling rate with
  associated duplication of resource usage, we wish to serve queries
  based on any sampling 
  rate from a single pass through the data.

\paragraph*{S3} \textsl{Control of User Database Access Rate:} We assume
  that in order to limit measurement access load on the database, the 
  access rate must be controlled to a desired level. This may occur 
through limiting either the frequency of exhaustive passes, or the
  frequency of individual accesses during the traversal.
\smallbreak

A challenge for creating a sampling regime satisfying these properties
 comes from certain  empirical facts: (i) the
distributions of some variables are markedly skewed, exhibiting heavy
tails, and (ii) there are strong correlations between some
pairs of variables (see Section~\ref{sec:data}). Uniform sampling is effective for
estimating ordinary distributions, but is a poor choice
for estimating the mass distribution of a heavy-tailed variable
\cite{Ch01}, since random omission from the sample of a single large
value can render estimates inaccurate. Instead, weighted
  sampling (see e.g., \cite{DLT07}) can be used to preferentially
  sample high-weight objects to provide accurate estimates of mass distributions.

\subsection{Contribution and Outline}

The contribution of this paper is first to specify a sampling
methodology that satisfies the properties [S1,2,3] above. 
We propose to meet the
condition [S1] by compiling, for each type of graph object (e.g., node,
link) a uniform sample and one or more weighted sample sets. This
raises two questions.
First: how should the weightings be chosen? 
A minimal way to do this is to determine the smallest set of (roughly)
independent heavy-tailed variables of interest, and use each of these
as a weight. In this approach, correlations between variables are
advantages in the sense that they may reduce the number of sample sets
that must be compiled. The second question is which sample set or
sets should be used in serving a given query. We propose
that the uniform sample be used for
ordinary distribution queries and queries concerning the mass
distribution of light-tailed variables, while a query concerning the
mass distribution of a heavy-tailed variable should use a weighted
sample set of either that variable, or, if absent, a sample set
weighted by a correlated heavy-tailed variable.

We propose to meet condition [S2] by using 
Priority Sampling\cite{DLT07} to compile, for each graph element type
and sample weighting, a \textsl{Master Sample} comprising a randomly ordered
subset of graph elements. A simple query on the master sample returns
the first $k$ items in order, with possible followup queries returning
the next $k$ and so on. A more complex query returns only elements that
match a given predicate. Here, the parameter $k$ determines the
effective sampling rate.

Priority Sampling can be implemented as either an offline or a streaming
algorithm. Correspondingly, condition [S3] can be achieved by limiting
a frequency of an offline redrawing of the sample sets or by
streaming graph elements into the sampler at some desired rate.

The outline of the paper is as follows. 
Section~\ref{sec:related} surveys the current state of related work
and positions our contribution with respect to this.
Section~\ref{sec:osn}
describes some classes of query used in  OSN
analysis and discuss the ramifications for observed distribution
features. Section~\ref{sec:method} describes
the compilation of master samples using Priority Sampling, and
procedures for estimating weight sums over subsets of population
elements specified by a selection predicate. We also show how to
estimate both ordinary and mass distributions from the samples. For
evaluation purposes we collected an OSN dataset described in
Section~\ref{sec:data}; the evaluations are described in
Section~\ref{sec:expt}. Our concluding discussion in
Section~\ref{sec:concl} sketches some possible extensions of our approach.

\section{A Survey of OSN Measurement}
\label{sec:related}

In this section we survey the current state of external and
internal measurement of OSN and node and link-level queries of common
interest. We describe the known effects of sampling on query accuracy
and the challenge that heavy-tailed distributions of the variable
bring to answering queries. We also explain how our work develops the state-of-the-art in distribution estimation and sample provisioning.

\subsection{External Measurements, Crawling and Random Walks}

There is a large literature on external measurements of OSNs that
illustrates the scale and complexity involved in data acquisition.
Early papers gathered data by crawling
Orkut~\cite{Mislove:2007:MAO:1298306.1298311} and
Twitter~\cite{wosn08,twit:huberman}.  A more recent, larger scale
study~\cite{moon-www10} used 20 machines simultaneously sending a
large number of requests to Twitter to fetch nearly 42 Million user
profiles and over 100 Million tweets. Another~\cite{cha-icwsm10}
fetched nearly 55 Million user profiles, 2 billion follow links, and
1.75 billion tweets.

The statistical properties of OSN data acquisition through crawling
has been examined in the framework of random walks on graphs. The
general theory of these goes back to \cite{erdos59debrecen}. 
A number of different sampling strategies for
graph traversal have been proposed, and evaluated for online networks
including the web, peer-to-peer, and social networks. These
evaluations have included examining the dependence of various
graphical statistics on the sampling rate. Unbiased sampling via
Metropolis-Hastings Random Walks was examined in
Facebook~\cite{gjoka:walking} and P2P
networks~\cite{Stutzbach:2006:USU:1177080.1177084}.  Forest Fire
Sampling \cite{1150479} explores the graph in a Markovian walk from
randomly selected nodes, while Frontier
Sampling~\cite{Ribeiro:2010:ESG:1879141.1879192} used multidimensional
random walks to mitigate trapping. Weighted random walks on graphs to
implement stratified sampling was proposed in \cite{Kurant:2011:WGM:1993744.1993773}. Graph sampling methods exploiting temporal clustering properties of OSN updates are presented in \cite{Ahmed:2010:TSS:1830252.1830253}.
Estimation of YouTube video counts through random prefix sampling was proposed
in \cite{Zhou:2011:CYV:2068816.2068851}. Estimating degree
distribution under network sampling is treated as a linear inversion problem 
by \cite{zhang2015}.

The effectiveness of different crawling and non-crawling sampling
strategies have been compared though their effects on estimation of
graphical statistics. Uniform sampling, BFS with
threshold and OPIC (online page importance) were compared in \cite{becchetti2006comparison}.
The effects of different sampling strategies
information diffusion metrics in Twitter were examined in ~\cite{choudhury10}. 
Sampling based estimates of the distributions of
popularity, length \& number of views reported in YouTube video
metadata were compared in \cite{Karkulahti2015}.
Our work is distinct from these in that we consider the
methodological underpinnings of what makes an effective sampling
strategy and its relation to the queries that are served.

\subsection{OSN Queries and Feature Distributions}

OSN research literature has focused on some popular questions about
various properties. These include characterization concerns such as
statistics about ``friendships"  (which in asymmetric OSNs like Twitter
includes notions of following and followers), outliers, connected and
disconnected components in the social graph, degree of separation,
homophily, assortativity,  and participation fraction \cite{becchetti2006comparison}. 
Dynamic properties such as reach, spread, and cascade
focus on users and applications that have higher influence, as
characterized by the speed
with which an application or a user's communication spreads
\cite{choudhury10}. This topic is of considerable commercial interest.

The joint distributions of graphical statistics in online networks are of great
interest for the user community \cite{sysomos} and have been
studied experimentally by a number of authors. The joint distributions
of OSN user characteristics (numbers of friends, followers, activity measures such as number of
posts) were studied in
\cite{twit:huberman}. \cite{Mislove:2007:MAO:1298306.1298311} found correlations between in
and out degrees in crawls of Flickr, Orkut and
YouTube. \cite{cha-icwsm10} studied the dependence between different
measures of influence in Twitter, namely, in-degree, retweets and
mentions. Our work is different from these in that we provide insight on
how best to sample based on the desired target statistics. Closer to
our approach is \cite{Wang:2013:DIF:2492517.2492662}, which proposed
Probability Proportional to Size sampling in OSN for estimation of
node degrees. Our work goes further: we consider the problem of how to estimate for
joint distributions of interest in applications. None of the above
works consider our problem of how to play out samples for analysis in
a optimal, tunable, and scaleable manner.

OSNs place various constraints on
the ability of users to obtain social graph data. Twitter provides
limited live samples via their Streaming API, while the Search API allows
queries against recent or popular tweets \cite{twitter-api}. Facebook
limits the number of API queries that can be submitted in a time period. Thus,
most of the research described here 
reports on a sample of the
data, which leads to questions about the nature and size of samples
needed to answer specific questions. 
The
evolution of this data, including the effective sampling rates, has
been studied over a multi-year period by  
\cite{liu-2014-twitter}, which also notes the inherent activity bias
of this datasets.
Even studies that report on full
crawls could be out of date after a relatively short period of time.

\subsection{The Challenge of Heavy Tails}

The approach of this paper builds on experience and methods from 
sampling Internet traffic flow records. The distribution of bytes per
flow is heavy-tailed \cite{FRC98}. Consequently, uniform sampling of flows, while
providing good estimates of counts of flows satisfying any predicate,
provides bad estimates of their byte counts because non-selection of
large flows greatly impact byte estimation accuracy. On the other
hand, weighting sampling by byte size enables accurate estimation of
bytes in flows satisfying any predicate, resulting in bad estimates of flow
counts. The conclusion here is that the heavy-tailed byte distribution
makes it difficult to simultaneously satisfy the accuracy requirements
of byte and flow estimation in a single sample. Instead, it is best to
compile two sample reservoirs, one with uniform sampling to serve
flow-level queries, one with bytes weighted sampling to serve
byte-level queries \cite{DL03}. In this paper we will exploit existing methods for
efficient Probability Proportional to Size sampling
\cite{DLT07,ADLT05}, but
the application of these to the problem of scaleable playout of
samples without repetition is new.

\section{Modeling OSN Queries}
\label{sec:osn}

Our work focuses on queries based on topological features of single
node or pairs of nodes, and on activity features.  
We consider the
class of queries that correspond to statements concerning the
(possibly joint) distribution of these node quantities, or some
summary statistic that integrates over the distributions. Thus, our
evaluation focuses on the question of how accurately the distributions
of these features can be estimated from the collection of sampling
reservoirs.  Many distributional queries and features of commercial
interest (see \cite{twit:huberman,sysomos}) can be abstracted into forms that
we now describe.

The set of single-user queries include:
\begin{itemize}
\item[(i)] \textsl{Distributions of single
features}, e.g., given $n$ find $x$ such that $x\%$ of users generate at least $n$ tweets each.
\item[(ii))]  \textsl{Self-weighted mass distributions of single
    features}, e.g., given $y$, find $x$ such that the top $x\%$ most active users generate $y\%$ of all tweets.
\item[(iii)] \textsl{Mass distributions of one feature weighted by
    another}, e.g., given $y$, find $x$ such that the top $x\%$ most followed users together generate $y\%$
 of all tweets.
\end{itemize}
In this paper we will focus on single use queries of the form just
described. However, the same methods can, in principle, be applied to
pair user queries including the following:
\begin{itemize}
\item[(iv)] \textsl{Joint distributions of features from two
users}, e.g., given $y$, find $x$ such that  $x\%$ of the total
activity between user pairs are between the top $y\%$ most active users.
\item[(v)] \textsl{Pairwise summary statistics such as assortativity}, i.e., the
  correlation between graph degrees of directly connected users.
\end{itemize}

\section{Samples, Playout \& Estimation}
\label{sec:method}

\def\km{k_{\mathrm{max}}}

This section describes the technical approach underpinning our
work. Section~\ref{sec:heavy} motivates weighted sampling as a
response to heavy-tailed distributions of node characteristics, while
Section~\ref{sec:thought} illustrates the ramifications of
correlations between different heavy-tailed
variables for sampling. Section~\ref{sec:priority} briefly reviews Priority
Sampling, while Section~\ref{sec:database} describes how it can be used
to fulfill the objectives S1--3 in the
introduction. Section~\ref{sec:distribution} shows how samples
selected through these means can be used to estimate the mass
distributions described in Section~\ref{sec:osn}.

\subsection{Heavy Tails and Weighted Sampling}\label{sec:heavy}

Our approach to sampling is guided by the principle
that sampling methodology should be chosen to match the statistical
characteristics of the data with the queries on that data.
In this case, the salient statistical features are:
\begin{itemize}
\item[(i)] Node features such as graph degree and activity that exhibit
  highly skewed, heavy-tailed
  distributions.
\item[(ii)] Node characteristics that exhibit varying amounts of statistical
  correlation; see \cite{twit:huberman} and Section~\ref{sec:evaluate}.
\end{itemize}
Uniform sampling estimates mass distributions poorly in case (i), since estimation
accuracy of large sizes
becomes highly sensitive to inclusion or exclusion from the
sample of large items, i.e., those for which the measure of interest,
such as node degree, is large. 
Weighted sampling reduces estimation variance by boosting the relative
selection probabilities of large items relative to small items.
The prime example is selecting a item of size $x$ with
Proportional to Size (PPS). A number of variants of this approach
exist, including weighted sampling without replacement,
Priority Sampling \cite{DLT07}, and Variance Optimal Sampling \cite{varopt}. All
these methods are able to construct a sample of a specified fixed
size. Priority Sampling and Variance Optimal sampling have efficient
implementations on data streams, and the latter minimizes estimation
variance compared with \text{any}  online or offline unbiased
estimator.

\subsection{Correlated Heavy Tails: A Thought Experiment}\label{sec:thought}

Consider two families $X=\{X_i,\ i=1,\ldots,n\}$ and $Y=\{Y_i,\
i=1,\ldots,n\}$ of heavy-tailed random weights, drawn independently
within each family. How do correlations
between $X$ and $Y$ affect PPS sampling?  We use the following
thought experiment to examine two extreme cases:
\begin{itemize}
\item[(i)] \textsl{Perfect
correlation:} $X_i=Y_i$ for all $i$;
\item[(ii)] \textsl{Independence:} $X_i$ and $Y_i$ are independent for
all $i$.
\end{itemize}
Under perfect correlations, PPS sampling using $X_i$ as weights
is effective for estimating large $Y_i$, since these equal the
corresponding $X_i$. But when $X$ and $Y$ are independent, the
occurrences of large values of $X$ and $Y$ are not correlated, so weighted sampling
based on $X$ will not select larger $Y$. 
In practice, correlations between node variables lie between these
extremes. Our experiments in Section~\ref{sec:evaluate} find varying degrees of
correlation between different heavy-tailed node variables. The less
strongly correlated variables are not interchangeable as weights for
PPS sampling.

\subsection{Priority Sampling}
\label{sec:priority}

The specific form of weighted sampling we use in the paper is Priority
Sampling \cite{DLT07}. From a population $\Omega$ of items with
weights $w_i$, Priority Sampling constructs a sample of any fixed size $k$ as follows.
First generate
for each $i$ a \textsl{priority} $\alpha_i=w_i/u_i$ where each $u_i$ is 
independently and uniformly distributed in $(0,1]$. Then 
$\Omega(k)$ comprises $k$ items of highest priority.  Define the
\textsl{threshold} $z(k)$ as the $(k+1)^{\textrm{st}}$ highest
priority. Then the effective sampling probability is is
$p_i(w_i,k)=\min\{1,w_i/z(k)\}$, and using the Horvitz-Thompson
inverse probability method \cite{HT52},
\be
\hat w_i = \left\{
\begin{array}{ll}
\max\{w_i,z(k)\}&\mbox{ if $i$ is sampled}\\
0 & \mbox{ if $i$ not sampled}\\
\end{array}
\right.
\ee
is an unbiased estimator of $w_i$ \cite{DLT07}. An unbiased
estimate of the subset sum $X(S)=\sum_{i\in X}x_i$ over any subset 
$S$ is just $\hat X(S)=\sum_{i\in S}\hat x_i$. A common case is when
$S$ is a set of items satisfying a predicate (e.g., users registered
in some region or with at least some number of relationships).

\subsection{Database Sampling and Playout}\label{sec:database}

In this section we show how Priority Sampling is well suited to serve
database queries based on any predicate with a mechanism possessing the following properties
\begin{itemize}
\item[(i)] Different sample sizes and predicates can be served from a
  master sample.
\item[(ii)] Multiple non-overlapping samples can be generated on the same predicate.
\item[(iii)] Sample volume or processing load can be controlled.
\end{itemize}
In each case, unbiased estimators can be constructed from the
corresponding samples.

\subsubsection{Creation of the Master Sample}
Following~\cite{ADLT05}, we first create the master sample: a descending priority-order sorted version $\Omega'$ of
the population $\Omega$ of database records, typically realized a sorted index into
the original set. This prodecure is performed 
one-time only: all randomness
occurs during this initial step. If the size of the master sample must be constrained to
a size $\km$, we take as $\Omega'$ the $\km$ elements of largest
priority.

\subsubsection{Samples of Given Size Over a Predicate}
A sample $S(k)$ of size $k$ over any predicate $S$ and $k$
can be constructed by selecting the first $k$ items in $\Omega'$ that
match the predicate. These are returned along with the sampling
weights $w_i$ and the sampling threshold $z(S,k)$ being the
$(k+1)^{\textrm st}$ largest priority of items matching $k$. If
$\Omega'$ is exhausted with only $1+k'$ matching items found, then the
first $k'$ of these are returned, along with $z(S,k')$.

\subsubsection{Non-overlapping Samples on a Given Predicate}
An initial sample $S(k)$ can be extended to size $k+j$ by adjoining
the next $j$ elements that match $S$, to yield $S(j+k)$. The original
$z(S,k)$ is discarded and $z(S,j+k)$ used for estimation
purposes. This step can be repeated as required; exhaustion of
$\Omega'$ is handled as above.

\subsubsection{Control of Computational Cost} Computational cost may
be controlled instead by selecting elements matching $S$ from within
the first $k$ of \textsl{all} elements of $\Omega'$. In this case, the
threshold reported is $z(k)$, the $(k+1)^{\textrm st}$ largest
priority in all of $\Omega'$.

\subsubsection{Estimation Accuracy} General bounds for estimation
accuracy of subset sums has been determined in \cite{Sze06}: an unbiased
estimate $\hat X$ of a subset sum $X$ based in $k$ samples obeys the
bound $\var(\hat X) \le \sqrt{X/(k-1)}$.

\begin{figure*}
\begin{center}
\begin{tabular}{cc}
\includegraphics[scale=0.66]{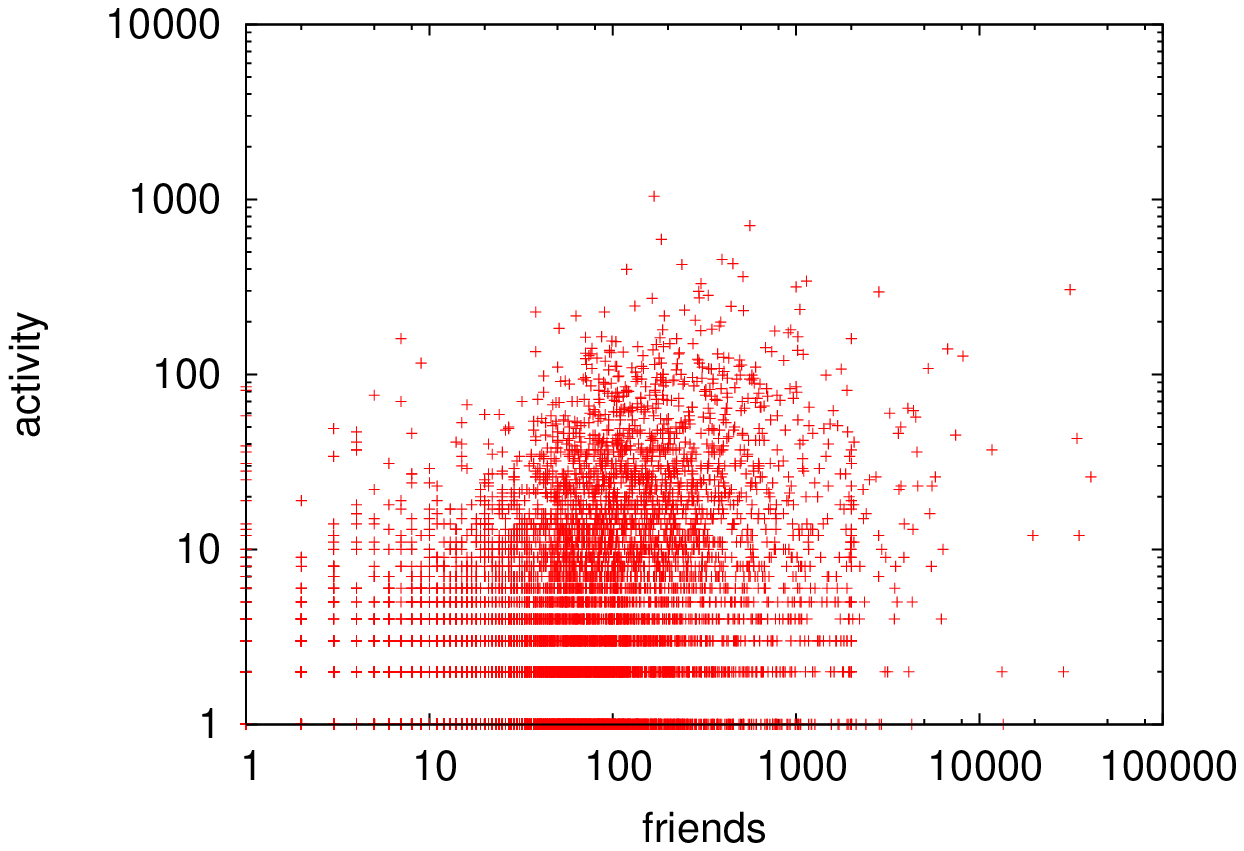}
&
\includegraphics[scale=0.66]{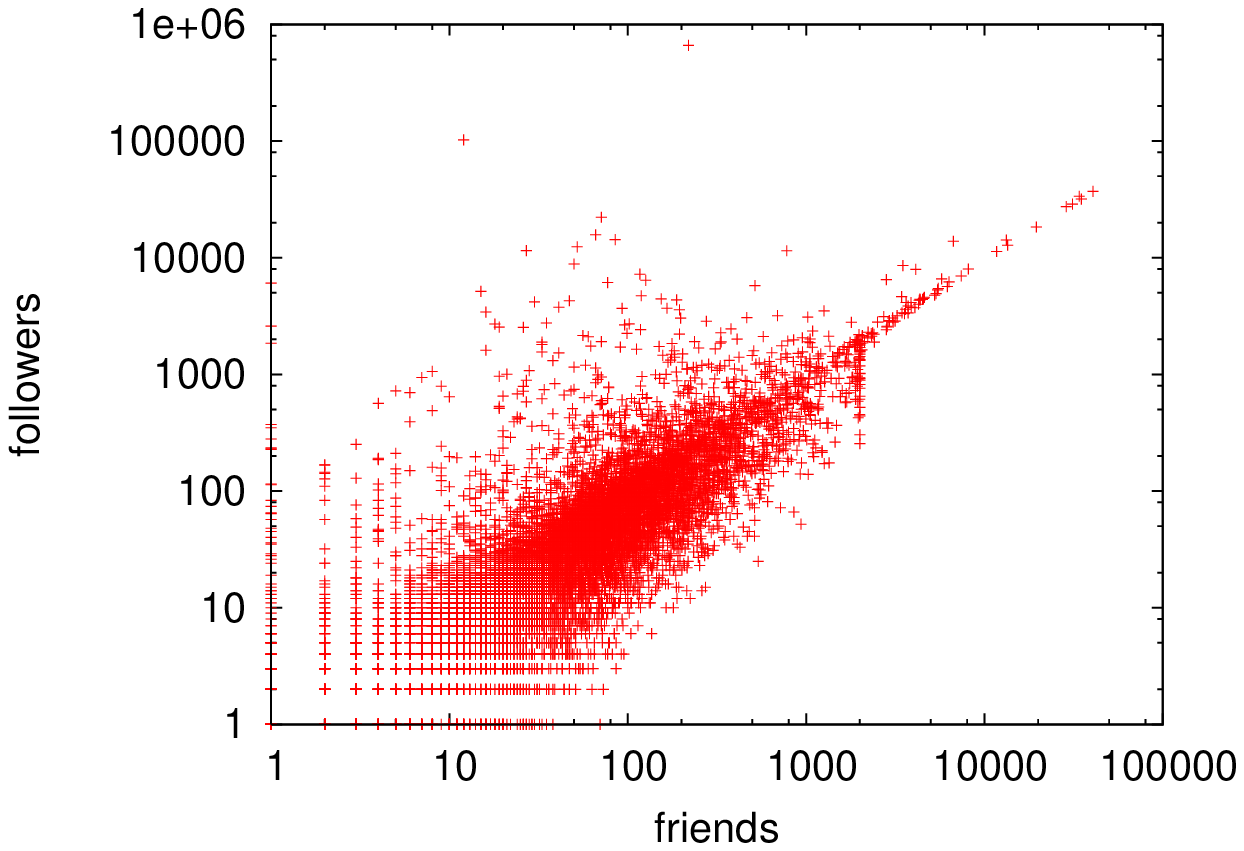}
\\
\end{tabular}
\caption{Pairwise scatter plots of node characteristics of 10,000
  random users.  
Left: weaker correlation, $r_s(\act,\fr)=0.44$. Right stronger
correlation $r_s(\fo,\fr)=0.82$ (Statistics from full dataset).}
\label{fig:sts:scatter}
\end{center}
\end{figure*}

\subsection{Distribution Estimation}\label{sec:distribution}

We now show how the joint distributions of user variables used in the
queries described in
Section~\ref{sec:osn} can be computed using subset sums whose
estimation from samples is
described above in Sections~\ref{sec:priority} and~\ref{sec:database}.
We model each user record in the database as containing a set of $m$
features $(x\up 1,\ldots,x\up m)$. A sampling weight is a value
$w=w(x\up 1,\ldots,x \up m)$ computed as a function of its
fields. These value may be directly reported in the record, and/or in
the simplest case be a single feature.

Given a predicate $\pi$, we will denote
\be\label{eq:xhat} \hat X\up
j(w,k,\pi)=\sum_{\pi(i)\wedge i\in\Omega(k)} x\up j_i /p_i(w_i,k) 
\ee
the estimate of the subpopulation sum of $x\up j$ over all records
satisfying the predicate $\pi$, based on a sample of size $k$ using
weights $w_i=w(x_i\up 1,\ldots,x_i\up m)$. We also distinguish 
the estimated subpopulation count:
\be\label{eq:nhat}
\hat N(w,k,\pi)=\sum_{\pi(i)\wedge i\in\Omega(k)} 1/p_i(w_i,k) 
\ee
We will omit the $\pi$ when all records are to be used, corresponding
to the case $\pi$ is identically true.

\begin{trivlist}
\item \textbf{Distribution of $X\up j$.} This estimate aims to yield
  statements of the form: \textsl{a proportion $q$ have records have
    $X\up j$ less than or equal to $y$.} 
Given a value $y$, define 
\be\label{eq:qy}
q(y) = \frac{\hat N\up j(w,k,x\up j \le  y)}{\hat N\up j(w,k)} 
\ee
Then $y$ estimates the $q(y)$-quantile of $X\up j$. To estimate the full CDF
of $X\up j$ we use the value $q(y)$ as $y$ is varied. We denote by
$\hat D_w(X\up j)$ the resulting distribution estimate.

\item \textbf{Mass distribution $X\up {\ell}$ by quantiles of $X\up j$.}
This estimate aims to
  yield statements of the form: \textsl{a proportion $r$ of the total
    mass of $X\up {\ell}$ is contained in a proportion $q$ of
    records with smallest $X\up j$.}
Given a value $y$, define 
\be\label{eq:mqy}
q(y) = \frac{\hat N(w,k,x\up j\le y)}{\hat N(w,k)},\quad
r(y) = \frac{\hat X\up \ell(w,k,x\up j \le y)}{\hat X\up \ell(w,k)}
\ee
Similar to above, $y$ estimates the $q(y)$-quantile of $X\up j$, while $r(y)$
is the proportion of the mass of $X\up {\ell}$ attributable to records with
$X\up j$ no greater than this quantile $y$. Varying $y$ as a parameter
yields the estimated curve $(q(y),r(y))$ of the mass distribution of $X\up
{\ell}$ by the quantiles of $X\up j$. $\hat M_w(X\up \ell,X\up
j)$ is the resulting mass distribution estimate.
\end{trivlist}

Note that in all the above examples, one can further restrict the
distribution to those items matching additional predicates that select
on the $(x\up 1,\ldots,x\up m)$.

\section{OSN Measurement Dataset}
\label{sec:data}

This section describes the acquisition and properties of the OSN dataset
used in the measurement-based study of this paper. 
Twitter supplies streamed data on a sampled basis via its Streaming
API interface \cite{twitter-api}, also referred to as the gardenhose.
This provided a sample of 5 public tweets out of every 100 (based
on the last two digits of the monotonically increasing status IDs),
written on a single TCP connection, significantly minimizing overhead
on Twitter servers. Twitter guaranteed randomness via their selection
process and their internal algorithm for assigning status IDs. Each
record streamed included nearly forty fields ranging from information
about the user who generated the tweet (including number of friends,
followers, status IDs) to information about the tweet itself (textual
content, language, etc.). During 2010, we gathered seven consecutive weeks of data
resulting in 75 million tweets and the associated information
generated by just over 8 million unique users. We selected 100,000
unique users using weighted sampling with each user weighted
proportionally to the number of tweets that the user originated in our
sample. Through the Twitter API we obtained the list of their friends and
followers\footnote{If friends/followers count exceeds 100, the API
limits the response to the 100 most recent}.

Our dataset is inherently
activity weighted since the more active users have a greater chance of
selection than the less active users; see \cite{liu-2014-twitter}. 
Thus, the
distributions of node variables in our data is different
from those in the unsampled user database
available to an OSN provider. However, this
does not change the essential thrust of our work (how best
to sample from correlated heavy-tailed distributions) since the
heavy-tails we observe are not created by sampling.

In graph language, each node corresponds to a user, and we denote by \fo\ and \fr\ the numbers of 
followers and friends as reported in the most recently observed status update
for each node. The activity \act\ is the number of status updates observed
for a given user.
Each of the 7.3 million direct graph links $(u_1,u_2)$ corresponds to
a follower relationship, i.e., $u_2$ is a follower of $u_1$.  An
example of a link feature combining features from both nodes is \ffan, the follower fanout of a link $(u_1,u_2)$, defined as
the ratio $\fo(u_2)/\fo(u_1)$. We believe this is an interesting
feature because high values of \ffan\ indicate links where a tweet
for user $u_1$ has the potential to be amplified if retweeted by user
$u_2$, since $u_2$ has a larger follower set.

As stated in Section~\ref{sec:intro}, the correlations amongst
different data features  have ramifications for the
choice of a set of different sample weightings. If a set of variables
is strongly mutually correlated, then a sample weighting based on any one of
them may be sufficient for estimating the mass distribution of any of
them. To assess the correlation amongst user features, we calculated
the Spearman rank correlation $r_s$ between each pair of features to
normalize the varying scaling behavior of the features. \fr\ and \fo\
are strongly correlated, with $r_s(\fr,\fo)=0.82$. Their correlation
with \act\ is weaker: $r_s(\act,\fo)=0.53$ and $r_s(\act,\fr)=0.44$.

We also use scatter plots of pairs of user features
to illuminate the relationships between the variables, using a uniform random sample of 10,000
users;  see
Figure~\ref{fig:sts:scatter}.
High
values of \fr\ and \act\ are relatively uncorrelated, while high values
of \fo\ and \fr\ are more strongly correlated. The
consequences of this behavior are discussed next.

\section{Measurement Study \& Evaluation}\label{sec:evaluate}\label{sec:expt}

We show how the estimation accuracy of both ordinary and mass
distributions depends both on the variables whose distribution is to
be estimated and the features used to weight sampling. To assess
accuracy we compare the estimated ordinary and mass distributions
$D_w(X)$ and $M_w(X,X')$ of $X$ (according to the quantiles of $X'$ in
the mass case), under the sample weighting $w$, with the true
distributions computed as ``estimates'' of all the data instead of
sampling. For node variables, we estimated the ordinary distributions
$\hat D_w(X)$ where $X\in\{\fo,\fr,\act\}$ and different weighting
schemes $w\in W=\{\uni,\fo,\fr,\act\}$ where \uni\ is uniform sampling
and other $w$ denote weighting by the specified node variable. The
mass distribution estimates were $\hat M_w(X,X')$ used
$X,X'\in\{\fo,\fr,\act\}$ and $w\in W$.  For the link variable
$\ffan$, we estimate the ordinary distribution using $\hat D_w(\ffan)$
with $w\in W'=\{\uni,\fo_1,\fo_2,\ffan\}$ where for $i\in\{1,2\}$,
$\fo_i$ is the $\fo$ value of node $u_i$ in a directed follower link
$(u_1,u_2)$. We estimated the mass distributions $M_w(\ffan,X')$ for
$X'\in \{\fo_1,\fo_2,\ffan\}$ and $w\in W'$.

Our principle comparison was between true and estimated distribution,
the difference characterized by the maximum absolute difference of the
cumulative distributions, similar to the Kolomogorov-Smirnov (KS) test
statistic \cite{Sachs84}.  For each $w,X$ (and $X'$ in the mass case) we
conducted 100 independent selections of 1000 samples, 
and summarized differences using the median KS statistic over the selections. 

\begin{table}
\begin{center}
\begin{tabular}{||c|ccc||}\hline
w & X=\fo & X =\fr & X = \act \\
\hline
\uni &0.066 &0.052 &0.352\\
\fr & 0.130 &0.152 &0.330\\
\fo & 0.116 &0.110 &0.346\\
\act &0.069 &0.055 &0.347\\
\hline
\end{tabular}
\caption{Ordinary Distribution Accuracy for Node Variables: Median KS
  statistic for $D_w(X)$ over 100 runs. KS noticeably smaller for
  weightings $w=\uni$ and $\act$}\label{tab:od:node}
\end{center}
\end{table}

\begin{table}
{
\begin{center}
\begin{tabular}{||c|c|ccc||}\hline
w & X'& X=\fo & X =\fr & X = \act \\
\hline
\uni & \fo &0.399 &0.144 &0.081\\
\uni &\fr &0.152 &0.141 &0.076\\
\uni &\act &0.172 &0.175 &0.083\\
\hline
\fo &\fo &0.024 &0.044 &0.089\\
\fo&\fr &0.024 &0.042 &0.088\\
\fo &\act &0.108 &0.162 &0.092\\
\hline
\fr &\fo & 0.291 &0.027 &0.075\\
\fr & \fr &0.196 &0.029 &0.086\\
\fr&\act &0.129 &0.165 &0.089\\
\hline
\act&\fo &0.370 &0.125 &0.026\\
\act&\fr &0.138 &0.135 &0.031\\
\act&\act &0.175 &0.169 &0.040\\
\hline
\end{tabular}
\caption{Mass Distribution Accuracy for Node Variables: Median KS statistic for
  $D_w(X,X')$ over 100 runs. Note KS generally smallest for
  weighting $w=X$.}\label{tab:md:node}
\end{center}
}
\end{table}

\begin{figure*}[t]
\begin{center}
\begin{tabular}{cc}
\epsfig{file=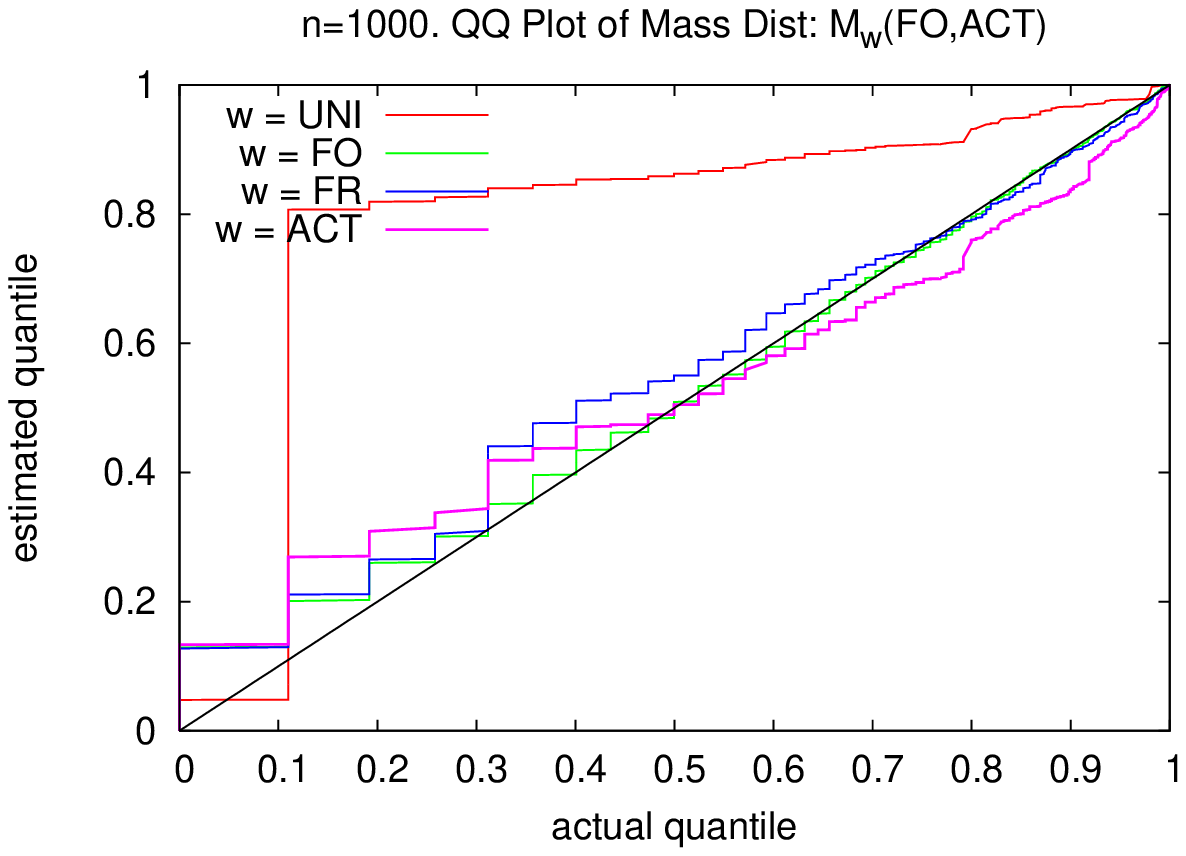,width=3.0in}
&
\epsfig{file=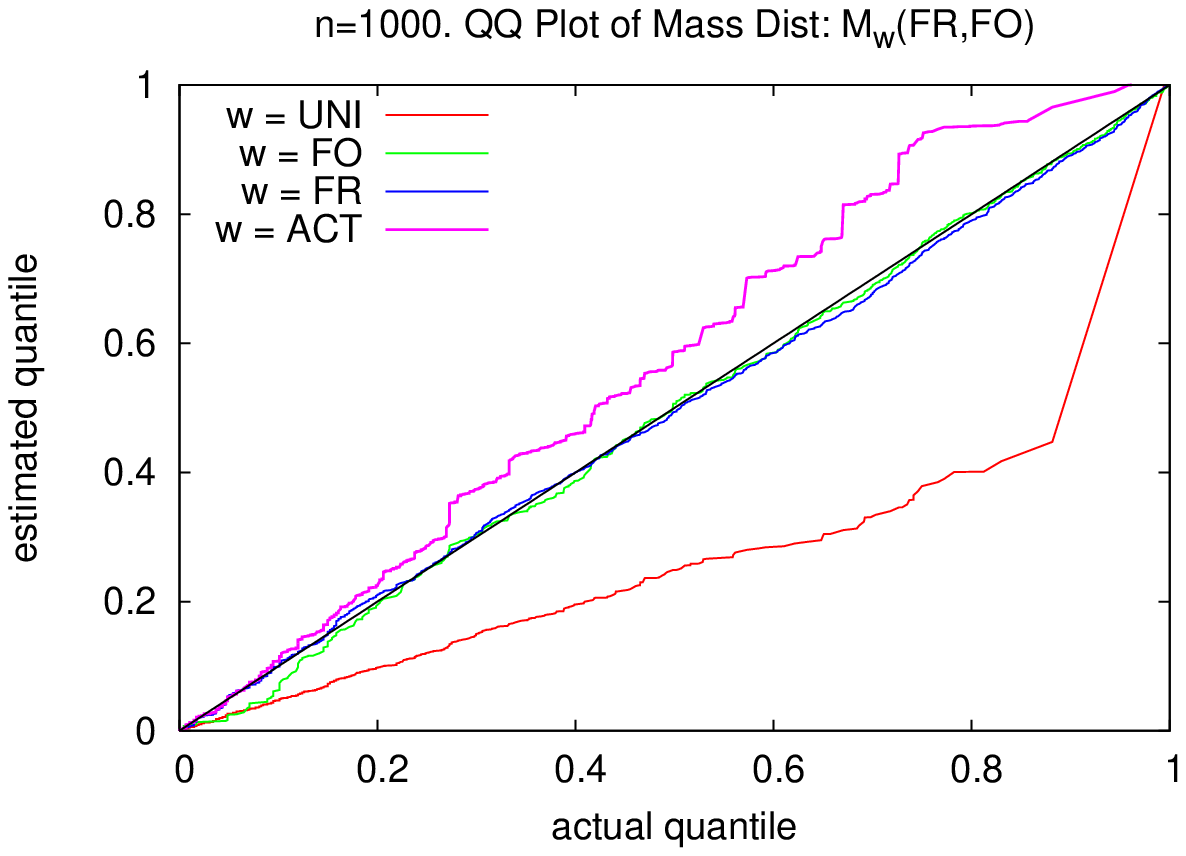,width=3.0in}
\\
\end{tabular}
\caption{Quantile-Quantile Plot of
  Estimate vs. Actual Mass Distribution from 1000 Samples of Node
  Variables. \fr\ and \fo\ best estimated using either as sample weights.}\label{fig:qq:node}
\end{center}
\end{figure*}

Results for node variables are shown in Tables~\ref{tab:od:node} and
~\ref{tab:md:node} for ordinary and mass distributions,
respectively. For ordinary distribution, the difference is noticeably
smaller for $\uni$ and $\act$ weightings than for $\fo$ and
$\fr$. This is unsurprising for $\uni$; for $\act$ it reflects the
absence of strong correlations between large values of $\act$ with
other variables. For the mass distribution, we observe the difference
for $D_w(X,X')$ is typically smallest when $w=X$, regardless of the
value of $X'$. In other words, the mass distribution of $X$ relative
to any other variable is generally most accurately estimated using $X$
as the sample weighting. However, the mass
distributions of $X=\fo,\fr$ have roughly equal estimation difference
whichever is used as the weighting $w$. This reflects the fact that
the distributions high values of $\fo$ and $\fr$ are fairly strongly
correlated; see Section~\ref{sec:data}.  Table~\ref{tab:md:link} shows results for the mass distribution of
the link variable \ffan. The smallest estimation error
using \ffan\ weighting, although $\fo_2$ weighting is reasonable,
performing noticeably better than with uniform of $\fo_1$ weighting.

\begin{figure*}[t]
\begin{center}
\begin{tabular}{cc}
\epsfig{file=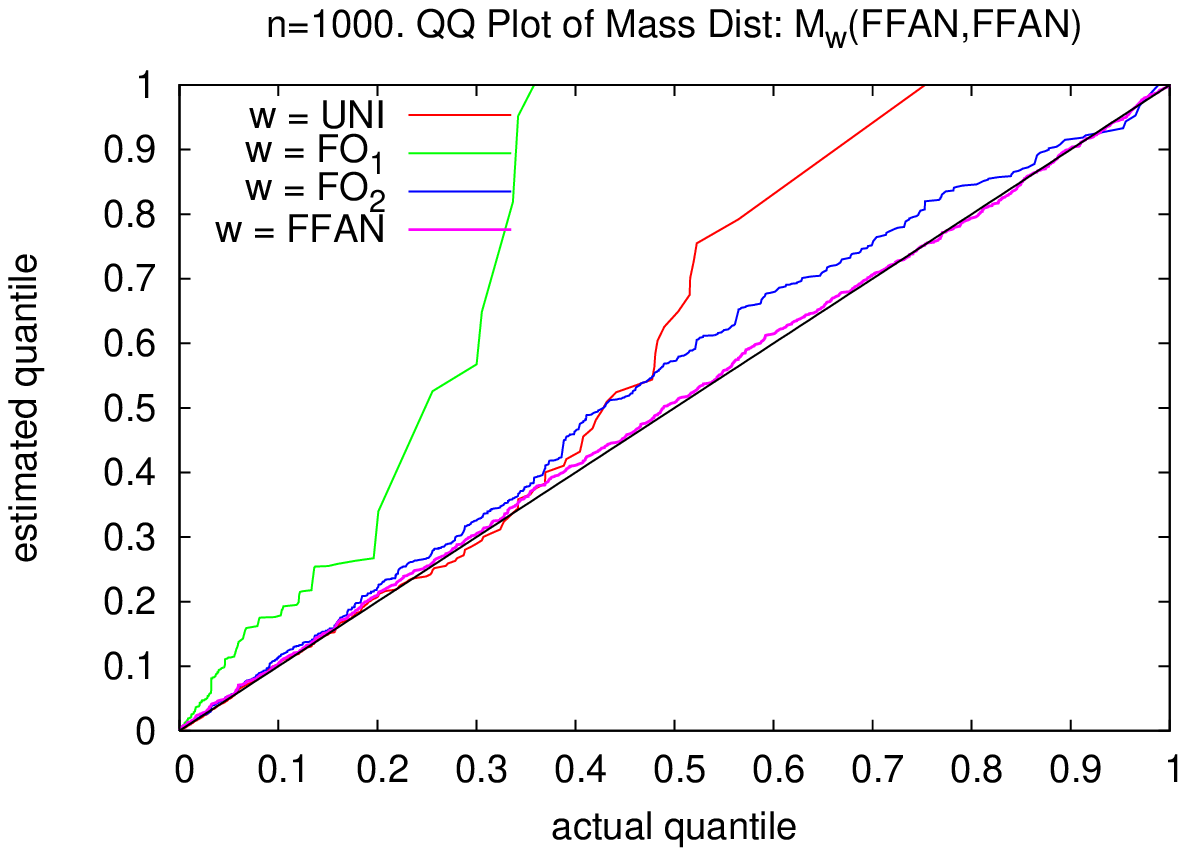,width=3.0in}
&
\epsfig{file=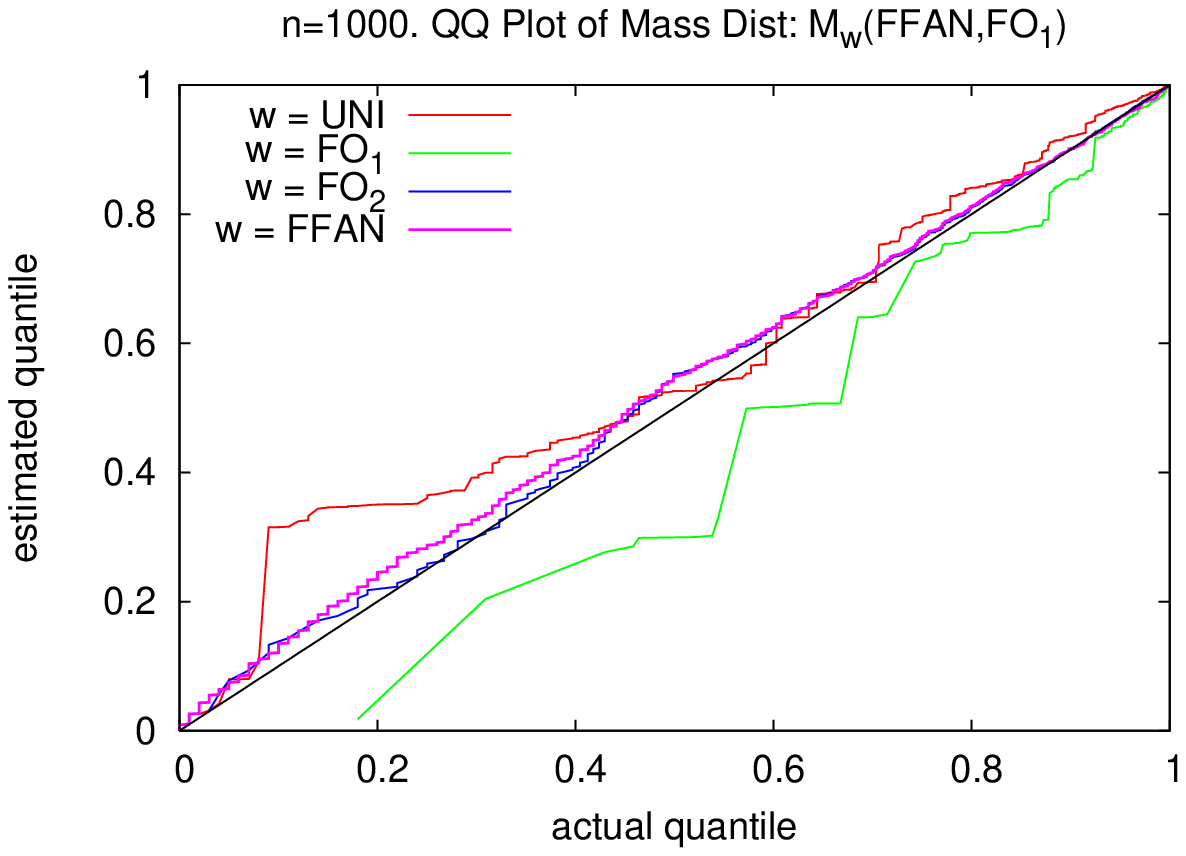,width=3.0in}
\\
\end{tabular}
\caption{Quantile-Quantile Plot of
  Estimate vs. Actual Mass Distribution from 1000 Samples of Link
  Variable \ffan. Best estimated using \ffan\ as weight, $\fo_2$
  reasonable, \uni\ and $\fo_1$ poor.}\label{fig:qq:link}
\end{center}
\end{figure*}

We illustrate a selection of
the estimated distributions through quantile-quantile plots of the
estimated vs. actual mass distributions for a section of variables.
Figure~\ref{fig:qq:node} shows the joint mass distributions of \fo\
weighted by \act\ (left) and \fr\ weighted by \fo\ (right), where each
line of the graph corresponds to a different sample weighing variable
(\uni, \fo, \fr, and \act). As expected, \uni\ has the poorest
performance, \fo\ and \fr\ the best (being highly correlated), while
\act\ has variable performance. Figure~\ref{fig:qq:link} displays
similar plots for the link fanout variable \ffan, for both the
self-weighted distribution (left) and the mass distribution with
respect to the initial node follower count \fo$_1$. In this case,
estimation accuracy is, unsurprisingly, best for sample weighting with
\ffan\ itself. However, sampling weighting with \fo$_2$, the numerator
of \ffan, also produces good results, while weighting with the
denominator \fo$_1$ is not so accurate. This is consistent with the
observation that large values of \ffan\ are driven more by large
values in \fo$_2$ than large values of \fo$_1$. 

\begin{table}
\begin{center}
\begin{tabular}{||cccc||}\hline
w = \uni& w = $\fo_1$ & w=$\fo_2$ & w=ffan\\ \hline
0.027&0.070&0.116&0.153\\
\hline
\end{tabular}
\caption{Ordinary Distribution Accuracy for Link Variable \ffan: Median KS
  statistic for $D_w(\ffan)$ over 100 runs.}\label{tab:od:link}
\end{center}
\end{table}

\begin{table}
\begin{center}
\begin{tabular}{||c|ccc||}\hline
w & X' =$\fo_1$& X' = $\fo_2$ & X'=\ffan \\
\hline
\uni & 0.231 &0.279 &0.300\\
$\fo_1$ &0.420 &0.462 &0.574\\
$\fo_2$ &0.065 &0.061 & 0.064\\
\ffan &0.034 &0.027 &0.025\\
\hline
\end{tabular}
\caption{Mass Distribution Accuracy for Link Variables \ffan: Median KS statistic for
  $D_w(\ffan,X')$ over 100 runs.}\label{tab:md:link}
\end{center}
\end{table}

\section{Discussion and Conclusions}\label{sec:concl}
This paper was motivated by commercial interest in 
the joint distributional properties of users and the connections
between them in  online social networks. Although much research has been devoted to
acquiring data through external crawling, OSN providers themelves are
well positioned both to access OSN data directly through their
customer databases and to monetize this as an information resource.

This paper proposes a method by which OSNs can provide samples of the
user graph of tunable size, in non-intersecting increments, with
sample selection that can be weighted to enhance accuracy when
estimating different features of the graph. A key experimental
conclusion what that neither uniform nor activity weighted sampling
were accurate for estimating mass distributions of friends,
followers and fanout, as compared with weighting by these features
themselves.

To bring our approach to fruition requires two steps which we identify
as future work. The first is to provide a method to
systematically determine the set of
required sample weightings as a function of the joint feature
distribution and the class of queries to be served. The second is to
provide an automated classification scheme that can dynamically
select the best weighted sample set or sets to serve a given query.

{\small
\bibliographystyle{acm}
\bibliography{paper,citations,fair,varopt,bibmaster}

\begin{thebibliography}{10}

\bibitem{Ahmed:2010:TSS:1830252.1830253}
{\sc Ahmed, N.~K., Berchmans, F., Neville, J., and Kompella, R.}
\newblock Time-based sampling of social network activity graphs.
\newblock In {\em MLG '10\/} (New York, NY, USA, 2010), ACM, pp.~1--9.

\bibitem{ADLT05}
{\sc Alon, N., Duffield, N., Lund, C., and Thorup, M.}
\newblock Estimating arbitrary subset sums with few probes.
\newblock In {\em Proc. 24th ACM Symp. on Principles of Database Systems
  (PODS)\/} (2005), pp.~317--325.

\bibitem{becchetti2006comparison}
{\sc Becchetti, {\mbox{L. et. al.}}.}
\newblock A comparison of sampling techniques for web characterization.
\newblock In {\em Workshop on Link Analysis ({LinkKDD})\/} (August 2006).

\bibitem{cha-icwsm10}
{\sc Cha, {\mbox{M. et al.}}.}
\newblock Measuring user influence in twitter: The million follower fallacy.
\newblock In {\em ICWSM\/} (2010).

\bibitem{Ch01}
{\sc Chaudhuri, {\mbox{S. et. al.}}.}
\newblock Overcomng limitations of sampling for aggregation queries.
\newblock In {\em ICDE'01\/} (2001), pp.~534--542.

\bibitem{choudhury10}
{\sc Choudhury, {\mbox{M. et. al.}}.}
\newblock How does the sampling strategy impact the discovery of information
  diffusion in social media.
\newblock In {\em ICWSM\/} (George Washington University, Washington, DC, May
  2010).

\bibitem{varopt}
{\sc Cohen, E., Duffield, N., Kaplan, H., Lund, C., and Thorup, M.}
\newblock Stream sampling for variance-optimal estimation of subset sums.
\newblock In {\em SODA '09: Proceedings of the twentieth Annual ACM-SIAM
  Symposium on Discrete Algorithms\/} (Philadelphia, PA, USA, 2009), Society
  for Industrial and Applied Mathematics, pp.~1255--1264.

\bibitem{DL03}
{\sc Duffield, N., and Lund, C.}
\newblock Predicting resource usage and estimation accuracy in an ip flow
  measurement collection infrastructure.
\newblock In {\em ACM SIGCOMM Internet Measurement Workshop\/} (2003).
\newblock Miami Beach, Fl, October 27-29, 2003.

\bibitem{DLT07}
{\sc Duffield, N., Lund, C., and Thorup, M.}
\newblock Priority sampling for estimation of arbitrary subset sums.
\newblock {\em J. ACM 54}, 6 (December, 2007), Article 32.
\newblock Announced at SIGMETRICS'04.

\bibitem{erdos59debrecen}
{\sc Erd{\"o}s, P., and R{\'e}nyi, A.}
\newblock {On Random Graphs}.
\newblock {\em Publ. Math. Debrecen 6\/} (1959), 290--297.

\bibitem{FRC98}
{\sc Feldmann, A., Rexford, J., and Caceres, R.}
\newblock Efficient policies for carrying web traffic over flow-switched
  networks.
\newblock {\em IEEE/ACM Transactions on Networking\/} (December 1998),
  673--685.

\bibitem{gjoka:walking}
{\sc Gjoka, {\mbox{M. et. al.}}.}
\newblock Walking in {Facebook}: A case study of unbiased sampling of {OSNs}.
\newblock In {\em INFOCOM\/} (March 2010), IEEE, pp.~1--9.

\bibitem{HT52}
{\sc Horvitz, D.~G., and Thompson, D.~J.}
\newblock A generalization of sampling without replacement from a finite
  universe.
\newblock {\em J. Amer. Stat. Assoc. 47}, 260 (1952), 663--685.

\bibitem{twit:huberman}
{\sc Huberman, B., Romero, D., and Wu, F.}
\newblock Social networks that matter: Twitter under the microscope.
\newblock In {\em Social Science Research Network\/} (December 2008).

\bibitem{Karkulahti2015}
{\sc Karkulahti, O., and Kangasharju, J.}
\newblock {\em Youtube Revisited: On the Importance of Correct Measurement
  Methodology}.
\newblock Springer International Publishing, Cham, 2015, pp.~17--30.

\bibitem{wosn08}
{\sc Krishnamurthy, B., Gill, P., and Arlitt, M.}
\newblock A few chirps about {Twitter}.
\newblock In {\em WOSN\/} (August 2008).

\bibitem{Kurant:2011:WGM:1993744.1993773}
{\sc Kurant, M., Gjoka, M., Butts, C.~T., and Markopoulou, A.}
\newblock Walking on a graph with a magnifying glass: stratified sampling via
  weighted random walks.
\newblock In {\em Proceedings of the ACM SIGMETRICS joint international
  conference on Measurement and modeling of computer systems\/} (New York, NY,
  USA, 2011), SIGMETRICS '11, ACM, pp.~281--292.

\bibitem{moon-www10}
{\sc Kwak, H., Lee, C., Park, H., and Moon, S.}
\newblock What is twitter, a social network or a news media?
\newblock In {\em WWW\/} (2010).

\bibitem{1150479}
{\sc Leskovec, J., and Faloutsos, C.}
\newblock Sampling from large graphs.
\newblock In {\em KDD\/} (New York, NY, USA, 2006), ACM, pp.~631--636.

\bibitem{Mislove:2007:MAO:1298306.1298311}
{\sc Mislove, A., Marcon, M., Gummadi, K.~P., Druschel, P., and Bhattacharjee,
  B.}
\newblock Measurement and analysis of online social networks.
\newblock In {\em Proceedings of the 7th ACM SIGCOMM Conference on Internet
  Measurement\/} (New York, NY, USA, 2007), IMC '07, ACM, pp.~29--42.

\bibitem{Ribeiro:2010:ESG:1879141.1879192}
{\sc Ribeiro, B., and Towsley, D.}
\newblock Estimating and sampling graphs with multidimensional random walks.
\newblock In {\em IMC '10\/} (New York, NY, USA, 2010), ACM, pp.~390--403.

\bibitem{Sachs84}
{\sc Sachs, L.}
\newblock {\em Applied Statistics}.
\newblock Springer, 1984.

\bibitem{Stutzbach:2006:USU:1177080.1177084}
{\sc Stutzbach, D., Rejaie, R., Duffield, N., Sen, S., and Willinger, W.}
\newblock On unbiased sampling for unstructured peer-to-peer networks.
\newblock In {\em Proceedings of the 6th ACM SIGCOMM Conference on Internet
  Measurement\/} (New York, NY, USA, 2006), IMC '06, ACM, pp.~27--40.

\bibitem{sysomos}
{\sc Sysomos}.
\newblock Insider twitter.
\newblock http://www.sysomos.com/insidetwitter, July 2009.

\bibitem{Sze06}
{\sc Szegedy, M.}
\newblock The {DLT} priority sampling is essentially optimal.
\newblock In {\em Proc. 38th STOC\/} (2006), pp.~150--158.

\bibitem{twitter-api}
{\sc Twitter}.
\newblock The streaming {API}'s.
\newblock https://dev.twitter.com/streaming/overview.

\bibitem{Wang:2013:DIF:2492517.2492662}
{\sc Wang, H., and Lu, J.}
\newblock Detect inflated follower numbers in osn using star sampling.
\newblock In {\em Proceedings of the 2013 IEEE/ACM International Conference on
  Advances in Social Networks Analysis and Mining\/} (New York, NY, USA, 2013),
  ASONAM '13, ACM, pp.~127--133.

\bibitem{liu-2014-twitter}
{\sc Yabing~Liu, C. K.-S., and Mislove, A.}
\newblock {The tweets they are a-changin': Evolution of Twitter users and
  behavior}.
\newblock In {\em {Proceedings of the 8th International AAAI Conference on
  Weblogs and Social Media (ICWSM'14)}\/} (Ann Arbor, MI, June 2014).

\bibitem{zhang2015}
{\sc Zhang, Y., Kolaczyk, E.~D., and Spencer, B.~D.}
\newblock Estimating network degree distributions under sampling: An inverse
  problem, with applications to monitoring social media networks.
\newblock {\em Ann. Appl. Stat. 9}, 1 (03 2015), 166--199.

\bibitem{Zhou:2011:CYV:2068816.2068851}
{\sc Zhou, J., Li, Y., Adhikari, V.~K., and Zhang, Z.-L.}
\newblock Counting youtube videos via random prefix sampling.
\newblock In {\em Proceedings of the 2011 ACM SIGCOMM Conference on Internet
  Measurement Conference\/} (New York, NY, USA, 2011), IMC '11, ACM,
  pp.~371--380.

\end{thebibliography}
}
\end{document}